# Time-resolved detection of spin-transfer-driven ferromagnetic resonance and spin torque measurement in magnetic tunnel junctions

Chen Wang[1], Yong-Tao Cui[1], Jordan A. Katine[2], Robert A. Buhrman[1], and Daniel C. Ralph[1,3*]

[1]Cornell University, Ithaca, New York 14853, USA
[2]Hitachi Global Storage Technologies, San Jose Res. Ctr., San Jose, CA 95135 USA
[3]Kavli Institute at Cornell, Cornell University, Ithaca, NY 14853, USA
* e-mail: ralph@ccmr.cornell.edu

**Abstract:**

Several experimental techniques have been introduced in recent years in attempts to measure spin transfer torque in magnetic tunnel junctions (MTJs). The dependence of spin torque on bias is important for understanding fundamental spin physics in magnetic devices and for applications.  However, previous techniques have provided only indirect measures of the torque and their results to date for the bias dependence are qualitatively and quantitatively inconsistent. Here we demonstrate that spin torque in MTJs can be measured directly by using time-domain techniques to detect resonant magnetic precession in response to an oscillating spin torque. The technique is accurate in the high-bias regime relevant for applications, and because it detects directly small-angle linear-response magnetic dynamics caused by spin torque it is relatively immune to artifacts affecting competing techniques. At high bias we find that the spin torque vector differs markedly from the simple lowest-order Taylor series approximations commonly assumed.

Spin transfer torque allows the magnetization in magnetic devices to be manipulated efficiently using the interaction of spin-polarized currents with ferromagnets.[1-3] The behavior of spin torque for high biases ($V$) applied to magnetic tunnel junctions (MTJs) provides a sensitive probe into the fundamental spin physics of hot electrons and is critical for applications, including next-generation memory devices and tunable oscillators.[4] However, measurements of spin torque in this regime have proven difficult. Previous approaches, based on different indirect measures of the torque, have yielded conflicting results.[5-17] Here we demonstrate an improved method of measuring spin torque by detecting, in a time-resolved fashion, the magnetic dynamics in linear response to the torque. Because this technique measures directly the small-angle precession caused by the torque, it is much less vulnerable than previous techniques to artifacts arising from heating or nonuniform magnetic dynamics. The method allows quantitative measurements of the bias dependence of the spin torque vector at large $|V|$, up to the breakdown voltage of the MTJ, and reveals behavior strikingly different from the approximations normally used to interpret experiments.

Several different approaches have been used previously in efforts to measure spin torque in MTJs. For measurements at low-to-moderate $|V|$, we believe that the most accurate technique is spin-transfer-driven ferromagnetic resonance (ST-FMR) with detection of magnetic precession via a dc mixing voltage.[5-9] However, this method fails at large $|V|$ due to an artifact associated with small changes in the dc resistance of MTJs in response to a microwave drive.[9] Analysis of the statistics of thermally-assisted switching can be used to extract the spin torque at large $|V|$,[10,11] but this method requires assuming a particular functional form for the bias dependence of the torque, it is sensitive to assumptions made about heating, and different analyses have yielded qualitatively different behaviors for the perpendicular, or "field-like," component of the spin torque vector.[10,11] Thermally-excited ferromagnetic resonance (TE-FMR)[13-17] is another technique that can in principle measure the perpendicular spin-torque component, based on the bias dependence of the precession frequency, but the results of this method can be questioned[9] because a bias may also affect the precession frequency via other mechanisms, including Joule[13] or Peltier heating or changes in the degree of spatially nonuniform dynamics due to lateral spin diffusion.[18,19]

In our technique, we apply a pulse of microwave current through the MTJ to apply an oscillating spin-transfer torque near the magnetic resonance frequency of one magnetic electrode, and we measure the resulting magnetic precession via oscillations of the MTJ resistance. Our measurement scheme is illustrated in Fig. 1a. We apply two electrical pulses simultaneously to the MTJ via a 50 Ω transmission line: a microwave-frequency (RF) pulse $V_{in}(t)$ with length 5-10 ns, long enough to reach steady-state resonant magnetic precession via the spin-torque effect, together with a longer square-wave pulse (~25 ns length) that starts a few nanoseconds earlier and ends several nanoseconds later than the RF pulse so that it provides the equivalent of a DC bias during the resonance measurement. We measure the signal reflected from the sample using a 12.5 GHz bandwidth sampling oscilloscope. The time-dependent part of the reflected voltage (prior to amplification) is:

$$V_{\text{ref}}(t) = \frac{(50\ \Omega)}{R_0 + (50\ \Omega)} I \Delta R(t) + \frac{R_0 - (50\ \Omega)}{R_0 + (50\ \Omega)} V_{in}(t). \qquad (1)$$

The first term on the right is the signal from the resistance oscillation that we aim to measure, with $I$ the effective DC current through the device provided by the square pulse and $\Delta R(t)$ the time dependent part of the MTJ resistance. The second term arises from the reflection of $V_{in}(t)$ from the impedance discontinuity between the 50 Ω cable and the sample with differential resistance $R_0$. One might consider trying to determine $\Delta R(t)$ by simply measuring the reflected signal during the time when $V_{in}(t)$ is nonzero, but the term due to the impedance discontinuity is generally 1-2 orders of magnitude larger than the term involving $\Delta R(t)$, and it is difficult to subtract this large background. Instead, we achieve a better signal-to-noise ratio by recording the reflected signal shortly (100 ps – 2 ns) after the falling edge of the RF pulse. In this time span, the resistance oscillation excited by the RF pulse (ST-FMR) is still present (although gradually decaying) while the strong background due to $V_{in}(t)$ is diminished. After subtracting the much-weakened background, we are able to clearly resolve the resistance oscillation of the MTJ. The details of the measurement are explained in the Methods section and Supplementary Note S1. A discussion about why the technique is relatively immune to artifacts from heating and spatially nonuniform magnetic dynamics is given in Supplementary Note S2.

We have measured MTJs with the following layer structure (in nm): bottom electrode [Ta(3)/CuN(41.8)/Ta(3)/CuN(41.8)/Ta(3)/Ru(3.1)], synthetic antiferromagnet (SAF) layer pinned to IrMn [IrMn(6.1)/CoFe(1.8)/Ru/CoFeB(2.0)], tunnel barrier [MgO$_x$], free layer [CoFe(0.5)/CoFeB(3.4)] and capping layer [Ru(6)/Ta(3)/Ru(4)]. Both the free layer and the SAF pinned layer are etched into a circular shape with diameter nominally either 80 nm or 90 nm. The devices have a nominal resistance-area (*RA*) product of 1.5 Ω-μm$^2$ and measured TMR ratios 85-100%. The capacitance is < 5 x 10$^{-14}$ F, small enough that it does not affect the experiment.[20] During the measurements, we apply a magnetic field *H* at an angle $\beta$ relative to the exchange bias of the SAF in order to produce a nonzero offset angle $\theta$ between the free layer and the reference layer of the SAF (Fig. 1b). We use the convention that positive bias corresponds to electron flow from the free layer to the SAF (favoring antiparallel alignment). We have measured ten samples, all with similar results, and we will report data from two of them. Sample 1 (nominally 90 x 90 nm$^2$) has a parallel-state resistance of 272 Ω and TMR of 91%, and sample 2 (nominally 80 x 80 nm$^2$) has a parallel-state resistance of 381 Ω and TMR of 97%.

Each time-domain measurement is performed in two steps. For each value of $\vec{H}$ for which a measurement is desired, we measure the differential resistance of the MTJ and also identify a different set of biasing conditions (field magnitude and direction) with the same value of differential resistance but which is non-resonant, in the sense that the magnetic resonance condition is well outside the frequency range of interest. We measure the RF pulse reflected from the sample under the non-resonant conditions to determine the background signal in the absence of magnetic dynamics. Fig. 2a shows a typical background signal near the falling edge of the RF pulse (vertical dashed line). We then measure the signal reflected from the sample for the biasing conditions that are of interest for measuring the magnetic dynamics, and subtract off the background (Supplementary Note S1 and Fig. S1). Figures 2b and 2c show the result of this subtraction for the case of *H* = 200 Oe, field direction $\beta$ = 90°, offset angle $\theta$ = 85°, *V* = 0.38 V for sample 1. Following the falling edge of the RF pulse, the measurement shows a resistance oscillation that decays gradually in time. The magnitude of oscillation corresponds to a maximum precession angle of about 1.5°, well within the linear-response regime. The decay rate for the oscillations agrees quantitatively with the magnetic damping rate measured by DC-detected ST-FMR in the same samples (Supplementary Note S4 and Fig. S2), indicating that the decay is due to a true decrease in precession amplitude, and is not dominated by dephasing between different repetitions of the measurement. Ideally our background subtraction should work not only after the RF pulse but also during the pulse, and we have indeed resolved the steady-state persistent resistance oscillation during the pulse as well (for time < 0 in Fig. 2b). This persistent oscillation confirms that the RF pulse is long enough to saturate the MTJ to steady-state dynamics. However, the oscillations measured during the pulse are noisy because of the large background that is subtracted, so we do not use them for quantitative analysis.

We fit the damped resistance oscillation following the falling edge of the RF pulse (Fig. 2c) to an exponentially decaying sinusoid $A_1 e^{-\Gamma t} \cos(\omega t + \Phi_1)$ with four parameters: amplitude $A_1$, phase $\Phi_1$, frequency $\omega$ and decay rate $\Gamma$, defining the centre of the falling edge of the pulse as time zero ($t = 0$). The fitting uncertainly for the phase is less than 0.04 radian, which corresponds to a time precision of ~ 1 ps. Using the same $t = 0$ point, we also fit the RF driving signal Fig. 2a prior to the falling edge of the pulse to a simple sinusoid $A_0 \cos(\omega_0 t + \Phi_0)$ with fitting parameters $A_0$ and $\Phi_0$. The phase of the magnetic response relative to the externally applied RF driving voltage is then $\Phi = \Phi_1 - \Phi_0$. In Fig. 2d we plot the normalized oscillation amplitude squared, $A_n^2 = (A_1/A_0)^2$, and relative phase $\Phi$ as a function of the RF pulse frequency for the same biasing conditions shown in the other panels of Fig. 2. As expected (see the Supplementary Note S3), the resonant response is accurately fit by a symmetric Lorentzian line shape and the phase changes by $\pi$ as the frequency is tuned through the resonance. We determine the natural frequency of the oscillator $\omega_m$ and the maximum normalized oscillation amplitude $A_{n,\max}$ from the fit to the amplitude response, and then determine the phase $\Phi_m$ of the magnetization precession at resonance by interpolating to the value at $\omega_m$ on a smooth polynomial fit of $\Phi$ vs. driving frequency.

From the values of $A_{n,\max}$, $\Phi_m$, and $\Gamma$ we make a quantitative determination of the spin torque vector. Because it is a torque, this vector is perpendicular to mean magnetization of the free layer and therefore we can specify it using two components, one in the plane determined by the magnetizations of the two electrodes ($\tau_\parallel$) and one perpendicular to this plane ($\tau_\perp$). A consideration of the equation of motion for the free layer magnetization shows that an oscillating in-plane torque on resonance gives rise to a resistance oscillation in phase with the driving voltage ($\Phi_m = 0$), while an oscillating perpendicular component contributes independently an out-of-phase resistance oscillation. Quantitatively, it follows that the voltage derivative of the torque, or "torkance"[21] $\partial \vec{\tau} / \partial V$, can be calculated from the in-phase and out-of-phase magnetic response at any bias $V$ (see Supplementary Note S3 for the derivation):

$$\left.\frac{\partial \tau_\parallel}{\partial V}\right|_\theta = \cos\Phi_m \frac{[R_0 - (50\Omega)][(R_0 + 50\Omega)]}{R_0(50\Omega)} \frac{\hbar}{2} \frac{M_S Vol}{\mu_B I} \left(\left.\frac{\partial R}{\partial \theta}\right|_I\right)^{-1} \Gamma A_{m,\max}, \quad (2)$$

$$\left.\frac{\partial \tau_\perp}{\partial V}\right|_\theta = -\sin\Phi_m \frac{[R_0 - (50\Omega)][(R_0 + 50\Omega)]}{R_0(50\Omega)} \frac{\hbar}{2} \frac{M_S Vol}{\mu_B I} \left(\left.\frac{\partial R}{\partial \theta}\right|_I\right)^{-1} \frac{\Gamma A_{m,\max}}{\Omega_\perp}. \quad (3)$$

Here $R_0$ is the measured differential resistance of the MTJ for the initial offset angle $\theta$ and bias $V$, $M_S Vol$ is the total magnetic moment of the free layer [estimated to be $1.8 \times 10^{-14}$ emu ($\pm 15\%$) for our $90 \times 90$ nm$^2$ devices and $1.6 \times 10^{-14}$ emu ($\pm 15\%$) for our $80 \times 80$ nm$^2$ devices based on vibrating sample magnetometer measurements of test films], $\mu_B$ is the Bohr magneton, $\partial R/\partial \theta|_I$ is the angular derivative of the DC resistance ($R = V/I$) of the MTJ, and $\Omega_\perp \approx 4\pi M_{eff} \gamma / \omega_m$ is a dimensionless factor (typically ~5 in our experiment), where $4\pi M_{eff} = 13 \pm 1$ kOe is the easy plane anisotropy strength of the free layer film (estimated by comparing our measured ST-FMR frequency to a Kittel formula[22]). For the data of Fig. 2, this analysis yields $\partial \tau_\parallel / \partial V = (0.44 \pm 0.10)(\hbar/2e)k\Omega^{-1}$, $\partial \tau_\perp / \partial V = (0.47 \pm 0.03)(\hbar/2e)k\Omega^{-1}$ at $V = 0.38$ V. The dominant experimental uncertainties are associated with determining the oscillation phase. By varying the amplitude of the square-wave pulse (and therefore DC bias), we can measure the spin transfer torkance for any value of |$V$| above about 0.1 V for our samples.

It should be noted that in response to a microwave excitation there is often more than one resonance mode in our devices at a given external field, usually one large amplitude mode and a second mode at least a factor of 3 smaller in amplitude. We suspect that the smaller mode involves oscillation of the magnetizations in the SAF, and that there may be coupling to the free layer oscillations. To limit the effects of mode coupling, for the data in this paper we have selected the direction and magnitude of the external magnetic field so that any secondary resonance mode is weak and well-separated in frequency from the primary resonance. The primary resonance can be identified with oscillations of the free magnetic layer based on the sign of the resonant response (increasing resistance for increasing bias) and the sign of the dependence of the magnetic damping on bias.

Figures 3a and 3b show our measurements of $\partial\tau_{//}/\partial V$ and $\partial\tau_{\perp}/\partial V$ over a large bias range for two samples. We display data up to $|V| = 0.6$ V because the distribution of breakdown voltages for our low-*RA* MTJs extends below 0.7 V. We have normalized the torkances by $\sin\theta$, because this is the angular dependence predicted for MTJs,[21,23] and indeed we find good agreement with this dependence within our experimental uncertainty. The figures show both the results of our new time-domain measurements for $|V| > 0.1$ V (square symbols), and the results in the same samples of the older DC-detected ST-FMR technique[7,9] for $|V| < 0.2$ V (triangles), which is a sufficiently small bias that this technique is reliable. In the range of overlap, $0.1$ V $< |V| < 0.2$ V, we find excellent quantitative agreement between these two independent techniques with no adjustment of parameters. This cross-check provides added confidence that both methods are quantitatively correct. By integrating the torkances with respect to voltage, we can plot the bias dependence of the spin-torque vector $\vec{\tau}(V)$ itself (Fig. 3c,d).

We observe that the in-plane component of the spin transfer torkance has an appreciable negative slope in the bias range $|V| < 0.2$ V in all of our MgO-based tunnel junctions, and is a factor of 3-4 stronger at high negative bias ($V < -0.2$ V) than at high positive bias ($V > 0.2$ V). Although this is a weaker bias dependence than had been suggested (incorrectly) in the past by uncorrected DC-detected ST-FMR measurements,[8,9] the in-plane component of the spin torque after integration does show significant nonlinearity and can become stronger by approximately a factor of 2.5 at large negative bias compared to positive bias (Fig. 3c,d). While an asymmetric bias dependence of the in-plane torkance is consistent with qualitative predictions[24,25] and *ab initio* calculations at low bias,[26] we suggest that a more quantitative theoretical understanding of the asymmetry at high bias should be a priority. Regarding the perpendicular component of the torkance, it had been known previously from calculations[24] and DC-detected ST-FMR measurements[7,8] that near $V = 0$ this component of the torkance in a symmetric MTJ has a linear dependence on bias (so that the perpendicular torque $\propto V^2$). We now observe departures from this behavior at high bias, in that $\partial \tau_\perp(V)/\partial V$ saturates (and $\tau_\perp(V)$ crosses over to an approximately linear dependence). Interestingly, the saturated value of perpendicular torkance differs significantly between positive and negative bias, which is forbidden by symmetry for an exactly symmetric MTJ when spin-flip scattering is negligible.[3] We suspect that this may be the result of a slight asymmetry in the structure of our MTJs (the bottom electrode is CoFeB grown on Ru, while the free layer is a CoFe/CoFeB bilayer grown on MgO) or the distribution of defects in the tunnel barrier or at the interfaces between the electrodes and the barrier. The strength of the perpendicular torque at the highest biases we measure is equivalent to a 30 Oe magnetic field, strong enough to play an important role in magnetic dynamics.

Our results have important consequences for interpreting many types of spin-torque experiments. Up to now, it has been assumed almost universally in analyzing experimental data that the bias dependence of the spin torque vector can be described by the simplest possible low-order Taylor series approximations, $\tau_\parallel \propto V$ and $\tau_\perp = a + bV + cV^2$ with *a*, *b,* and *c* constants. This is often done, for example in extrapolating from finite-temperature measurements to determine zero-temperature critical currents and activation energy barriers for switching,[27-30] and it is also the underlying assumption in analyzing the statistics of switching to determine the strength of the spin torque vector.[10,11] Our measurements indicate that these Taylor-series approximations can become seriously inaccurate at high bias, so that extrapolations based on these approximations should not be expected to yield quantitative results. Furthermore, analyses of asymmetric MTJs have generally assumed that the main effect of the asymmetry on the perpendicular torque is to add a linear bias dependence to $\tau_\perp(V)$, together with the quadratic dependence present for symmetric junctions.[11] The difference in the saturated values of $\partial \tau_\perp / \partial V$ for positive and negative *V* that we observe at high bias suggest that the effects of asymmetry may be significantly larger at high bias than at low bias, and may take functional forms different than can be expressed by a lowest-order Taylor approximation.

The asymmetry we observe for the in-plane spin torque may, in addition, help to explain the observation that the critical voltage for switching from the antiparallel (AP) to parallel (P) configuration (negative bias in our convention) is often lower than for P-to-AP switching.[8,11,31-34] However, we note that in the thermally-assisted switching regime this effect can be somewhat mitigated by contribution from the perpendicular spin torque, which always favors the AP state for our MTJs.

In summary, we have introduced a time-resolved experimental technique that can provide direct observations of small-angle magnetic precession in MTJs in response to spin transfer torque. We have used the technique to obtain quantitative measurements of the amplitude and phase of precession excited by a resonant microwave pulse, from which we determine both the in-plane and perpendicular components of the spin torque vector at high bias. These results show that the lowest-order Taylor approximations for bias dependence of the spin torque vector, employed almost universally, can be markedly inaccurate at high bias. We expect that the same time-resolved measurement technique will also be able to provide new insights about a wide range of other interesting phenomena in MTJs, including nonlinear magnetic dynamics in response to large spin torques,[35] phase locking of magnetic auto-oscillations to microwave inputs,[36,37] time-resolved coupling between magnetic modes, and spin torques in very low RA MTJs for which pinholes may contribute new effects.[12,38]

**Methods**:

The microwave (RF) pulse in our measurement is generated by using a mixer and a short gating pulse to modulate a continuous-wave source. The peak-to-peak voltage produced by the RF pulse at the MTJ is 20-40 mV, the pulse length is 5-10 ns, and the rise and fall times of the RF pulse are approximately 100 ns and 180 ns respectively. The RF pulse is combined with the longer (25 ns) square wave pulse via a power combiner. Employing a square pulse instead of an actual DC voltage to bias the MTJ significantly reduces the probability of electrical breakdown at high voltages and allows us to access higher biases than in DC-dectected ST-FMR or TE-FMR experiments. A sinusoidal reference signal generated by one RF source is used to synchronize the clock of the oscilloscope and is also used (after frequency division) to trigger both pulse generators and the oscilloscope at 100 kHz. These synchronization measures fix the phase of the RF pulse and allow for signal averaging. We generally average for 25 seconds (about 1250 repetitions for 2000 sampling points).

Calibration procedures and the methods used to measure the differential resistance $R_0$ and the DC resistance $R$ at high biases are described in the Supplementary Note S5.

**Acknowledgements**:

We thank D. Mauri of Hitachi Global Storage Technologies (now at Western Digital Corp.) for providing junction thin film stacks that we used to fabricate the tunnel junctions. Cornell acknowledges support from ARO, ONR, DARPA, NSF (DMR-1010768), and the NSF/NSEC program through the Cornell Center for Nanoscale Systems. We also acknowledge NSF support through use of the Cornell Nanofabrication Facility/NNIN and the Cornell Center for Materials Research facilities.

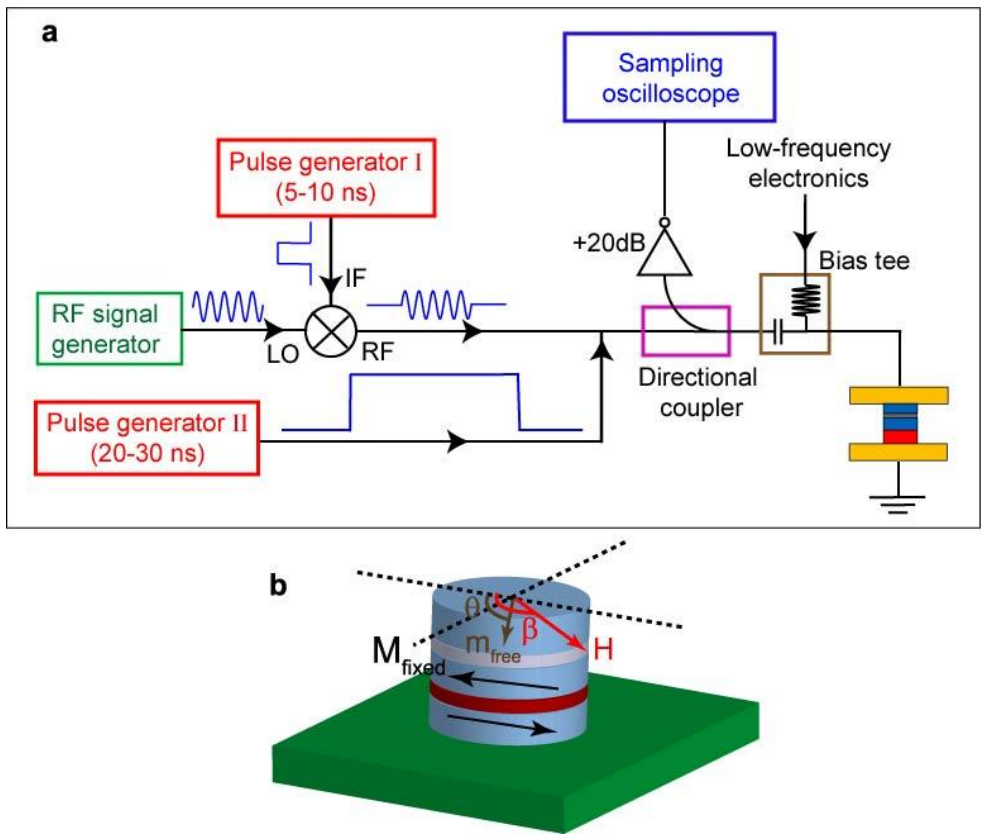

Fig. 1. **Diagrams of the experiment. a**, Schematic of our measurement circuit. **b**, Geometry of our MTJ devices.

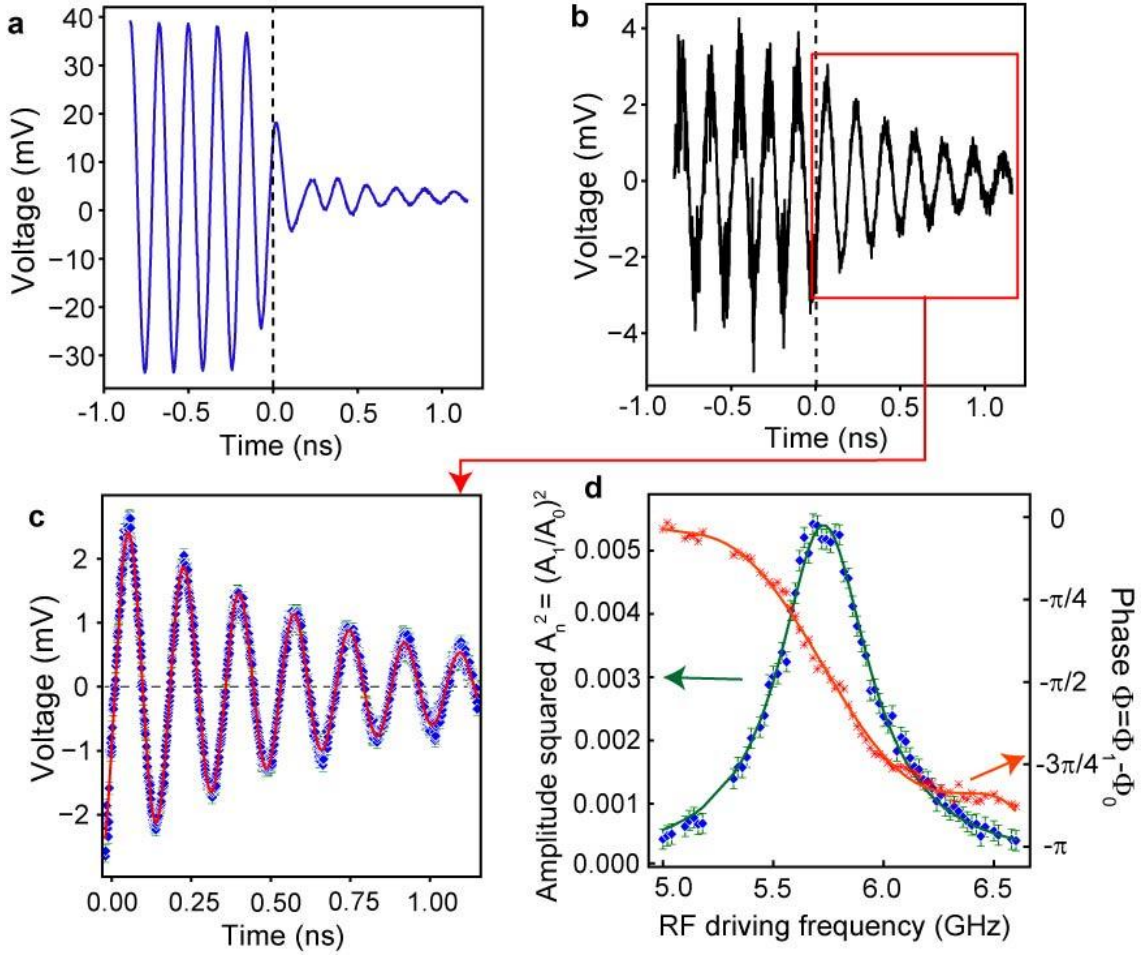

Fig. 2. **Time-resolved measurements of spin-torque-driven magnetic resonance**. **a**, Falling edge of the applied waveform (for a 5.8 GHz RF pulse), measured on reflection from sample 1 with the MTJ biased at a non-resonant state with magnetic field $H = 600$ Oe (field direction $\beta = 94°$) and voltage $V = 0.38$ V. This waveform represents the background in the resonance measurement. **b**, The reflected voltage waveform from sample 1 near magnetic resonance after background subtraction for an applied magnetic field $H = 200$ Oe, field direction $\beta = 90°$, offset angle $\theta = 85°$, $V = 0.38$ V. This signal is proportional to the resistance oscillations of the MTJ excited by the resonant RF pulse shown in **a**. **c**, Close-up view of the measured oscillations in **b** from the time span after the falling edge of the RF pulse, along with a fit to a decaying sinusoidal curve (red line). **d**, Dependence on the frequency of the RF drive pulse (for the same biasing conditions as in **b**) for the normalized oscillation amplitude squared $A_n^2 = (A_1/A_0)^2$ and the oscillation phase relative to the drive pulse, $\Phi = \Phi_1 - \Phi_2$. The green curve is a symmetric Lorentzian fit to the amplitude data, and the red curve is a smooth polynomial fit of the phase.

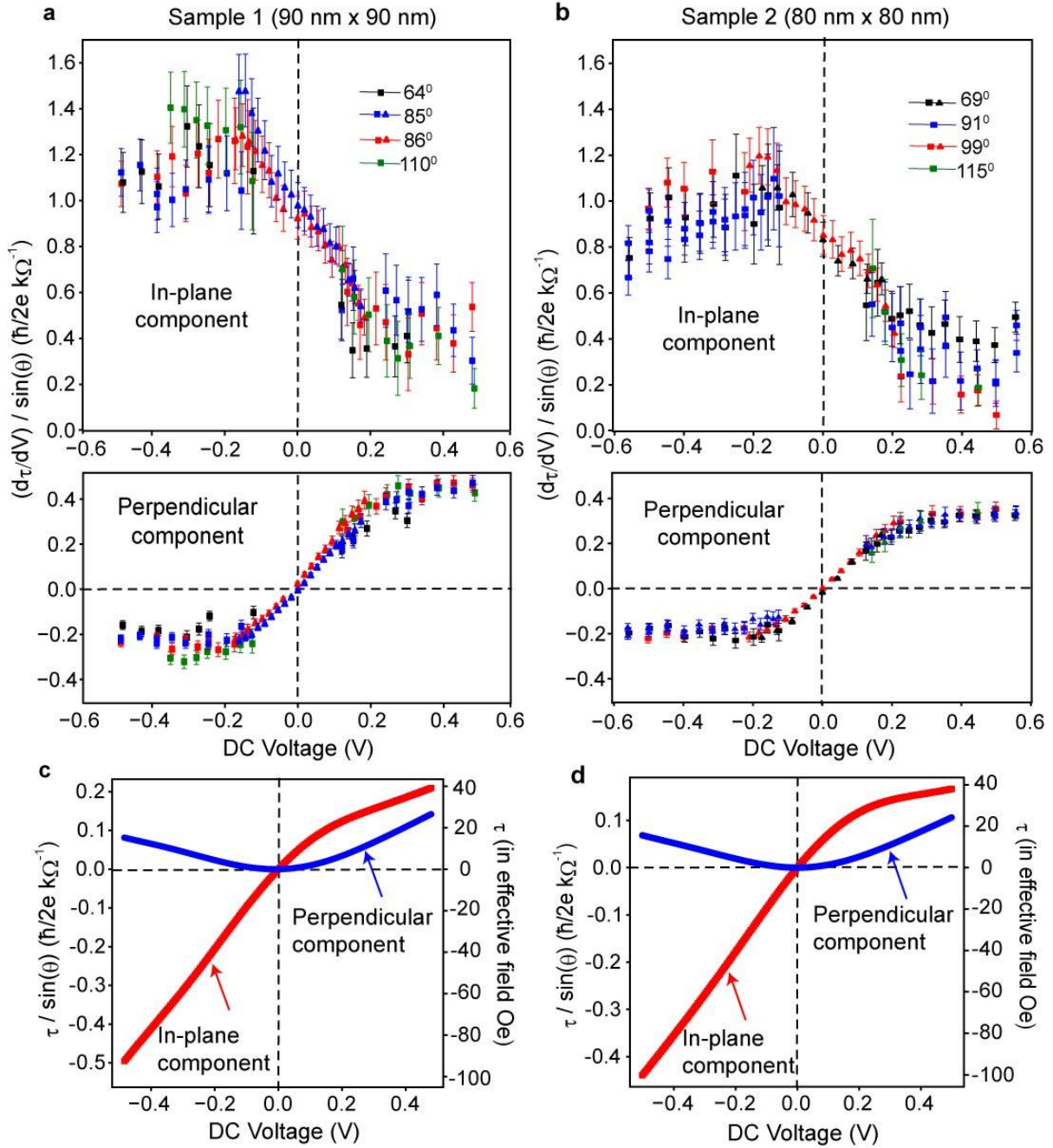

Fig. 3. **Measured bias dependence of the spin-transfer torque vector**. **a,b**, In-plane and perpendicular components of the torkance vector $\partial\vec{\tau}/\partial V$ (normalized by $\sin\theta$) as a function of bias voltage for **a** sample 1 and **b** sample 2, for different initial offset angles, $\theta$. The square symbols correspond to our time-resolved measurements and the triangles to DC-detected ST-FMR on the same samples. **c,d**, In-plane and perpendicular components of the spin transfer torque $\vec{\tau}$ (normalized by $\sin\theta$) for **c** sample 1 and **d** sample 2, determined by integrating the data in **a,b** after averaging over the different initial offset angles.

## Supplementary Material

**Supplementary Note 1: Data acquisition and background subtraction in the time-domain ST-FMR measurement**

In order to separate the resistance oscillation signal from the residual background oscillation of the reflected RF pulse, for any given state of the device (for magnetic field magnitude $H$, field direction $\beta$ and bias voltage $V$), we identify a non-resonance reference state of the same device with equal impedance to perform background measurements. With the same impedance, the reflected signal of the same RF pulse should be the same for both states, so that the difference between the two waveforms represents the net signal produced by the resistance oscillation.

To find a reference state for the ($H$, $\beta$, $V$) state, we first determine the differential resistance $R_0$ of the device at the ($H$, $\beta$, 0) state with a low-frequency (about 1 kHz) lock-in measurement, then we increase the external magnetic field to $H'$ (typically by 500 Oe) to shift the resonance peak completely out of the frequency range of our measurement, and finally we fine-tune the field direction to $\beta'$ (typically by a few degrees) so that the lock-in measurement of the resistance $R_0(H', \beta', 0)$ is equal to $R_0(H, \beta, 0)$. Throughout the measurement under the same field set ($H$, $\beta$) at arbitrary biases $V$, we use the ($H'$, $\beta'$, V) state as the reference state for the ($H$, $\beta$, $V$) state. Because the resistance is strongly dependent on bias voltage in MTJs, $R_0(H, \beta, V)$ is very different from $R_0(H, \beta, 0)$. However, if we assume that an applied bias voltage does not change the direction of the magnetization dramatically (we estimate the effect of the static spin torque can result in a change of $\theta$ by at most 3° for all our field conditions), to a good approximation $R_0 (H', \beta', V) = R_0 (H, \beta, V)$ should still hold because the two states have the same offset angle and bias voltage. We verified this by comparing the reflected waveform of an off-resonance RF pulse (typically with a frequency lower than the resonance of either state) at the two states, and found that the difference between the two was negligible compared to our measured signal after subtraction.

To make a time-domain measurement of the resistance oscillation at the state ($H$, $\beta$, $V$) under a given RF driving pulse, we first record the waveform at the state of interest ($H$, $\beta$, $V$) after averaging for 25 seconds (about 1250 repetitions for 2000 sampling points) (the red curve in Fig. S1a,b). We then keep the applied RF driving pulse on to maintain the same phase and amplitude, switch quickly to the reference state (*H'*, *β'*, *V*), and record the background waveform, again averaging for 25 seconds (the black curve in Fig. S1a,b). Afterwards we switch back to the original state (*H*, *β*, *V*) and repeat the measurement (the blue curve in Fig. S1a,b which is hardly visible since it overlaps with the red curve). It is important to record the the signal trace both before and after recording the background trace because either the falling edge of the RF pulse or the phase of the RF pulse can drift with respect to the time base of the oscilloscope by a few picoseconds over several minutes of measurements. We discard any data taken when this drift is significant (*i.e.* more than one picosecond) during the overall measurement time of about 1.5 minutes. Finally we average the signal traces and subtract the background, giving a final result such as shown in Figures 2b and 2c in the main text. The resulting waveform is proportional to the resistance oscillation of the MTJ.

**Supplementary Note 2: Comments about the effects of heating and spatially non-uniform magnetic dynamics.**

We claim in the main paper that ST-FMR with time-domain detection of the resonant magnetic dynamics is much less vulnerable than competing techniques to artifacts arising from heating or nonuniform magnetic dynamics. We will expand on those issues here.

Our technique has little sensitivity to heating because, at least for a circular sample, an oscillatory temperature should not drive magnetic precession with a definite phase relationship relative to the input RF signal. Therefore, the presence of heating should not give rise to any artifact signals in a time-domain experiment in which the measurements are made following the end of the drive pulse. For a non-circular sample and an equilibrium free-layer orientation away from symmetry direction for the magnetic anisotropy, it is possible for oscillatory heating to drive precession via modulation of the magnetic anisotropy fields, but the torques we measure show reasonable agreement with the expected angular dependence ($\propto \sin\theta$) without this effect, which indicates that it is not a significant factor. Heating in our experiment might lead to a decrease in the average moment of the free layer in our samples at high bias, therefore increasing slightly the apparent value of the torkances for large |*V*|. However, we anticipate that this is a small effect unless the bias is high enough that the effective magnetic temperature approaches the Curie temperature (> 800 K).

One may also question the applicability of the macrospin approximation we use to extract the value of the spin-torque vector from our measurements. Our measurements are designed to minimize the likelihood of any significant spatial nonuniformities in the magnetic dynamics, so that the macrospin model should be a good approximation. We employ small, circular samples (as small as 80 nm diameter) with an in-plane equilibrium magnetic state, take measurements in an external in-plane magnetic field of 200-400 Oe that promotes a uniform initial state, and excite only small angle precession from this state (~1° mean deflection). However, even in the presence of small deviations from the macrospin approximation, the results of ST-FMR with time-domain detection should be largely insensitive to these deviations, particularly relative to previous techniques for determining the spin torque vector. Our measurements begin with a large equilibrium offset angle between the magnetic orientation of the reference layer and the average orientation of the free layer, so that the spin torque applied throughout the free layer is large, in the same direction, and approximately uniform. By measuring resistance, we determine, to a good approximation, the average deflection induced by the spin torque over the entire free layer. Via the time-domain measurement, we also measure directly the oscillation decay rate integrated over the whole sample. Under these circumstances, average precession amplitudes we measure, and hence the values of the spin-torque vector that comprise our final result, should be determined by conservation of spin angular momentum integrated over the whole free layer (total spin angular momentum absorbed in each cycle = spin angular momentum lost to damping in each cycle), even if the magnetic dynamics may contain small spatial nonuniformities.

**Supplementary Note 3: Determination of the spin transfer torque vector from the ST-FMR oscillation signal**

In this derivation we assume that the dynamics of the magnetic free layer can be approximated by a simple macrospin model, and we calculate the magnetic response to an oscillating spin torque. We have commented above in Supplementary Note 2 about the possible consequences of micromagnetic effects beyond the macrospin model. Our calculation builds on the derivation of the ST-FMR signal presented in the appendix of ref. 9, with one correction and one addition. The correction is that here we take into account that the directions of the in-plane and out-of-plane components of the spin torque vector will rotate slightly relative to the sample axes as the free layer undergoes small-angle precession. The failure to take this into account in our previous paper led to a factor-of-two error in calculating the bias dependence of the effective magnetic damping, and the correction brings our results into accord with the result in references 13 and 14. The addition is that we self-consistently take into account a feedback effect,[39] that when a DC bias is applied to a precessing magnetic tunnel junction, the resistance oscillations arising from the precession will modify the high-frequency part of the bias across the junction. This effect can become significant under conditions of large DC bias and small effective magnetic damping.

We assume that the magnetic moment of the free layer has a constant magnitude $M_s Vol$ and we denote its instantaneous direction by the unit vector $\hat{m}(t)$. We define $\hat{z}$ as the equilibrium direction of the free layer (and assume that it is within the plane of the thin-film sample to a good approximation even at non-zero bias $V$), $\hat{x}$ as the direction perpendicular to the film plane, and $\hat{y}$ is the in-plane direction satisfying $\hat{z} = \hat{x} \times \hat{y}$. At the equilibrium position, $\hat{m} = (0,0,1)$. Small-angle precession of the free layer in response to a DC + oscillatory voltage $V(t) = V + \delta V(t) = V + \text{Re}(\delta V e^{i\omega t})$ can be expressed in the form $\hat{m}(t) = (m_x(t), m_y(t), 1)$ or $\hat{m}(t) = \text{Re}(m_x e^{i\omega t}, m_y e^{i\omega t}, 1)$, where $|m_x|, |m_y| << 1$. We assume that the reference magnetic layer remains fixed in the in-plane direction $\hat{M} = (0, \sin\theta, \cos\theta)$. Because of the large magnetic anisotropy of the thin film sample, $|m_x| << |m_y|$ and therefore the oscillations of the angle $\theta$ between the two magnetizations during precession can be expressed to a good approximation as $\delta\theta(t) = -\text{Re}(m_y e^{i\omega t})$.

Within the macrospin approximation, the dynamics of the free layer moment are governed by the Landau-Lifschitz-Gilbert-Slonczewski (LLGS) equation:

$$\frac{d\hat{m}}{dt} = -\gamma \hat{m} \times \vec{H}_{\text{eff}} + \alpha \hat{m} \times \frac{d\hat{m}}{dt} + \gamma \frac{\tau_\parallel(V,\theta)}{M_s Vol} \hat{m} \times \frac{\hat{m} \times \hat{M}}{|\hat{m} \times \hat{M}|} + \gamma \frac{\tau_\perp(V,\theta)}{M_s Vol} \frac{\hat{m} \times \hat{M}}{|\hat{m} \times \hat{M}|}, \quad \text{(S1)}$$

where $\tau_\parallel(V,\theta)$ and $\tau_\perp(V,\theta)$ are in-plane (Slonczewski) component and out-of-plane (field-like) components of the spin torque, written as a function of voltage and offset angle between the two layers. (Note that this equation is corrected from the form of the LLGS equation used in ref. 9 by taking into account that the in-plane and out-of-plane unit vectors will shift direction as $\hat{m}(t)$ precesses.) $\vec{H}_{\text{eff}} = -N_x M_{\text{eff}} m_x \hat{x} - N_y M_{\text{eff}} m_y \hat{y}$ is the effective field acting on the free layer, with $N_x = 4\pi + H/M_{eff} \approx 4\pi$ since the in-plane field we apply (200-400 Oe) is much weaker than the easy-plane anisotropy of the film (~13 kOe). $N_y \approx H/M_{eff}$, neglecting anisotropy and interlayer coupling. $\gamma = 2\mu_B/\hbar$ is the absolute value of the gyromagnetic ratio, and $\alpha$ is the Gilbert damping coefficient. For small RF excitation voltages, the magnitude of the spin-torque components can be Taylor-expanded,

$$\tau_\|(V+\delta V(t), \theta+\delta\theta(t)) = \tau_\|(V,\theta) + \frac{\partial \tau_\|}{\partial V}\delta V(t) + \frac{\partial \tau_\|}{\partial \theta}\delta\theta(t)$$
$$\tau_\perp(V+\delta V(t), \theta+\delta\theta(t)) = \tau_\perp(V,\theta) + \frac{\partial \tau_\perp}{\partial V}\delta V(t) + \frac{\partial \tau_\perp}{\partial \theta}\delta\theta(t). \tag{S2}$$

As noted above, the directions of the torque components also change directions slightly during precession,

$$\hat{m} \times \frac{\hat{m}\times\hat{M}}{|\hat{m}\times\hat{M}|} = -\hat{y} + m_x \cot\theta\hat{x} + m_y\hat{z}, \quad \frac{\hat{m}\times\hat{M}}{|\hat{m}\times\hat{M}|} = -\hat{x} - m_x\cot\theta\hat{y} + m_x\hat{z}, \tag{S3}$$

to first order in $m_x$ and $m_y$. After substituting Eq. (S2) and Eq. (S3) into Eq. (S1), the time-dependent part is, to first order,

$$\frac{dm_x(t)}{dt} = -\gamma m_y(t) N_y M_{eff} - \alpha\frac{dm_y(t)}{dt} + \frac{\gamma}{M_S Vol}\tau_\| m_x(t)\cot\theta - \frac{\gamma}{M_S Vol}\left[\frac{\partial \tau_\perp}{\partial V}\delta V(t) - \frac{\partial \tau_\perp}{\partial \theta}m_y(t)\right],$$

$$\frac{dm_y(t)}{dt} = \gamma m_x(t) N_x M_{eff} + \alpha\frac{dm_x(t)}{dt} - \frac{\gamma}{M_S Vol}\left[\frac{\partial \tau_\|}{\partial V}\delta V(t) - \frac{\partial \tau_\|}{\partial \theta}m_y(t)\right] - \frac{\gamma}{M_S Vol}\tau_\perp m_x(t)\cot\theta. \tag{S4}$$

Using the notation $m_x(t) = \text{Re}(m_x e^{i\omega t})$, $m_y(t) = \text{Re}(m_y e^{i\omega t})$ and $\delta V(t) = \text{Re}(\delta V e^{i\omega t})$,

$$i\omega m_x = -m_y(\gamma N_y M_{eff} + i\alpha\omega) + \frac{\gamma}{M_S Vol}\tau_\| m_x\cot\theta - \frac{\gamma}{M_S Vol}\left[\frac{\partial \tau_\perp}{\partial V}\bigg|_\theta \delta V - \frac{\partial \tau_\perp}{\partial \theta}\bigg|_V m_y\right]$$

$$i\omega m_y = m_x(\gamma N_x M_{eff} + i\alpha\omega) - \frac{\gamma}{M_S Vol}\left(\frac{\partial \tau_\|}{\partial V}\bigg|_\theta \delta V - \frac{\partial \tau_\|}{\partial \theta}\bigg|_V m_y\right) - \frac{\gamma}{M_S Vol}\tau_\perp m_x\cot\theta. \tag{S5}$$

In our experiments, we apply an external RF voltage $V_{in}(t) = V_{in}\text{Re}(e^{i\omega t})$ to the sample, which is related to the true voltage and current at the sample by the following microwave circuit equations,

$$V(t) = V + \text{Re}(\delta V e^{i\omega t}) = V + V_{in}\text{Re}(e^{i\omega t}) + \text{Re}(V_{ref}e^{i\omega t})$$
$$I(t) = I + \text{Re}(\delta I e^{i\omega t}) = I + V_{in}\text{Re}(e^{i\omega t})/(50\ \Omega) - \text{Re}(V_{ref}e^{i\omega t})/(50\ \Omega) \tag{S6}$$

Also, the resistance of the MTJ imposes the restrictions,

$$V(t) = V + \left.\frac{\partial V}{\partial I}\right|_\theta \delta I(t) + \left.\frac{\partial V}{\partial \theta}\right|_I \delta\theta(t) = V + R_0 \delta I(t) + I\left.\frac{\partial R}{\partial \theta}\right|_I \delta\theta(t) \tag{S7}$$

where $R_0 = \left.\partial V/\partial I\right|_\theta$ is the differential resistance of the MTJ, and $R = V/I$ is the DC resistance of the MTJ. From Eq. (S6) and Eq. (S7), together with the relation $\delta\theta(t) = \text{Re}(-m_y e^{i\omega t})$,

$$V_{in} + V_{ref} = R_0 \frac{V_{in} - V_{ref}}{(50\,\Omega)} - I\left.\frac{\partial R}{\partial \theta}\right|_I m_y. \tag{S8}$$

Therefore,

$$V_{ref} = \frac{R_0 - (50\,\Omega)}{R_0 + (50\,\Omega)} V_{in} - \frac{(50\,\Omega)}{R_0 + (50\,\Omega)} I\left.\frac{\partial R}{\partial \theta}\right|_I m_y \tag{S9}$$

and

$$\delta V = V_{in} + V_{ref} = \frac{2R_0}{R_0 + (50\,\Omega)} V_{in} - \frac{(50\,\Omega)}{R_0 + (50\,\Omega)} I\left.\frac{\partial R}{\partial \theta}\right|_I m_y. \tag{S10}$$

By expressing the magnetic equations of motion in terms of the applied RF drive $V_{in}$ (which does not change due to feedback when the free layer rotates) rather than the actual oscillating voltage across the tunnel junction $\delta V$ (which does change due to feedback), we can account for the feedback self consistently. Therefore we substitute Eq. (S10) into Eq. (S5):

$$i\omega m_x = -m_y\left(\gamma N_y M_{eff} + i\alpha\omega\right)$$
$$- \frac{\gamma}{M_S Vol}\left[\frac{2R_0}{R_0 + (50\,\Omega)}\left.\frac{\partial \tau_\perp}{\partial V}\right|_\theta V_{in} - \left(\left.\frac{\partial \tau_\perp}{\partial \theta}\right|_V + \frac{(50\,\Omega)}{R_0 + (50\,\Omega)} I\left.\frac{\partial R}{\partial \theta}\right|_I \left.\frac{\partial \tau_\perp}{\partial V}\right|_\theta\right) m_y - \tau_\parallel^0 m_x \cot\theta\right]$$
$$i\omega m_y = m_x\left(\gamma N_x M_{eff} + i\alpha\omega\right)$$
$$- \frac{\gamma}{M_S Vol}\left[\frac{2R_0}{R_0 + (50\,\Omega)}\left.\frac{\partial \tau_\parallel}{\partial V}\right|_\theta V_{in} - \left(\left.\frac{\partial \tau_\parallel}{\partial \theta}\right|_V + \frac{(50\,\Omega)}{R_0 + (50\,\Omega)} I\left.\frac{\partial R}{\partial \theta}\right|_I \left.\frac{\partial \tau_\parallel}{\partial V}\right|_\theta\right) m_y + \tau_\perp^0 m_x \cot\theta\right]$$
$$\tag{S11}$$

Solving Eq. (S11) for $m_y$ yields

$$m_y = \frac{R_0}{R_0 + (50\,\Omega)} \frac{\gamma V_{in}}{M_S Vol} \frac{2\omega}{(\omega^2 - \omega_m^2 - i2\omega\Gamma_0)}\left[i\left.\frac{\partial \tau_\parallel}{\partial V}\right|_\theta + \frac{\gamma N_x M_{eff}}{\omega}\left.\frac{\partial \tau_\perp}{\partial V}\right|_\theta\right], \tag{S12}$$

which close to resonance $\omega \approx \omega_m$ becomes

$$m_y = \frac{R_0}{R_0 + (50\,\Omega)} \frac{\gamma V_{in}}{M_S Vol} \frac{1}{\omega - \omega_m - i\Gamma_0}\left[i\left.\frac{\partial \tau_\parallel}{\partial V}\right|_\theta + \frac{\gamma N_x M_{eff}}{\omega_m}\left.\frac{\partial \tau_\perp}{\partial V}\right|_\theta\right], \tag{S13}$$

where

$$\omega_m \approx \gamma M_{eff} \sqrt{N_x \left[ N_y - \frac{1}{M_{eff} M_S Vol} \left( \frac{\partial \tau_\perp}{\partial \theta} \bigg|_V + \frac{(50\,\Omega)}{R_0 + (50\,\Omega)} I \frac{\partial R}{\partial \theta} \bigg|_I \frac{\partial \tau_\perp}{\partial \mathcal{V}} \bigg|_\theta \right) \right]} \tag{S14}$$

$$\Gamma_0 \approx \frac{\alpha \gamma M_{eff}(N_x + N_y)}{2} - \frac{\gamma}{2 M_S Vol} \left( \tau_\parallel \cot\theta + \frac{\partial \tau_\parallel}{\partial \theta} \bigg|_V + \frac{(50\,\Omega)}{R_0 + (50\,\Omega)} I \frac{\partial R}{\partial \theta} \bigg|_I \frac{\partial \tau_\parallel}{\partial \mathcal{V}} \bigg|_\theta \right). \tag{S15}$$

Substituting Eq. (S13) in Eq. (S9), we have for the reflected signal

$$V_{ref} = \frac{R_0 - (50\,\Omega)}{R_0 + (50\,\Omega)} V_{in}$$
$$- \frac{(50\,\Omega)}{R_0 + (50\,\Omega)} \left( \frac{R_0}{R_0 + (50\,\Omega)} \right) I \frac{\partial R}{\partial \theta} \bigg|_I \frac{\gamma}{M_S Vol} \frac{1}{(\omega - \omega_m - i\Gamma_0)} \left[ i \frac{\partial \tau_\parallel}{\partial \mathcal{V}} \bigg|_\theta + \frac{\gamma N_x M_{eff}}{\omega_m} \frac{\partial \tau_\perp}{\partial \mathcal{V}} \bigg|_\theta \right] V_{in}, \tag{S16}$$

where the first term is the measured background $V_{bg}$ in our experiment, and the second term is the measured signal arising from the resistance oscillation $V_{sig}$. Therefore the normalized complex precession amplitude $Ae^{i\Phi}$ that we analyze in our time-domain experiment takes the form

$$Ae^{i\Phi} = V_{sig}/V_{bg} = -\frac{R_0(50\,\Omega)}{[R_0 - (50\,\Omega)][R_0 + (50\,\Omega)]} I \frac{\partial R}{\partial \theta} \bigg|_I \frac{\gamma}{M_S Vol} \frac{1}{(\omega - \omega_m - i\Gamma_0)} \left[ i \frac{\partial \tau_\parallel}{\partial \mathcal{V}} \bigg|_\theta + \frac{\gamma N_x M_{eff}}{\omega_m} \frac{\partial \tau_\perp}{\partial \mathcal{V}} \bigg|_\theta \right]. \tag{S17}$$

This is a convenient quantity for analysis because amplifier gains and transmission losses cancel on account of the normalization. At the resonance frequency,

$$A_{n,\max} e^{i\Phi_m} = \frac{R_0(50\,\Omega)}{[R_0 - (50\,\Omega)][R_0 + (50\,\Omega)]} I \frac{\partial R}{\partial \theta} \bigg|_I \frac{\gamma}{M_S Vol\, \Gamma_0} \left[ \frac{\partial \tau_\parallel}{\partial \mathcal{V}} \bigg|_\theta - i \frac{\gamma N_x M_{eff}}{\omega_m} \frac{\partial \tau_\perp}{\partial \mathcal{V}} \bigg|_\theta \right]. \tag{S18}$$

From this, and using that $\gamma = 2\mu_B/\hbar$, we have the final expression we use to evaluate the time-domain results [Equations (2) and (3) in the paper].

$$\frac{\partial \tau_\parallel}{\partial \mathcal{V}} \bigg|_\theta = \cos\Phi_m \frac{[R_0 - (50\Omega)][(R_0 + 50\Omega)]}{R_0(50\Omega)} \frac{\hbar}{2} \frac{M_S Vol}{\mu_B I} \left( \frac{\partial R}{\partial \theta} \bigg|_I \right)^{-1} \Gamma_0 A_{m,\max}, \tag{S19}$$

$$\frac{\partial \tau_\perp}{\partial \mathcal{V}} \bigg|_\theta = -\sin\Phi_m \frac{[R_0 - (50\Omega)][(R_0 + 50\Omega)]}{R_0(50\Omega)} \frac{\hbar}{2} \frac{M_S Vol}{\mu_B I} \left( \frac{\partial R}{\partial \theta} \bigg|_I \right)^{-1} \frac{\Gamma_0 A_{m,\max}}{\Omega_\perp}. \tag{S20}$$

Here $\Omega_\perp \approx 4\pi M_{eff} \gamma / \omega_m$. We have only to verify that the quantity $\Gamma_0$ introduced above as the half width at half maximum of the FMR lineshape has the same value as the decay rate $\Gamma$ of the damped resistance oscillations in the time-domain experiment.

To prove this, we analyze the equation of motion in the case that $V_{in}(t) = 0$, in the presence of a DC current. Noting that $\delta V(t) = \frac{(50\ \Omega)}{R_0 + (50\ \Omega)} I \frac{\partial R}{\partial \theta}\bigg|_I m_y(t)$ from Eq. (S10) and substituting $m_x(t) = \text{Re}(m_x e^{(-\Gamma + i\omega)t})$, $m_y(t) = \text{Re}(m_y e^{(-\Gamma + i\omega)t})$ and $\delta V(t) = \text{Re}(\delta V e^{(-\Gamma + i\omega)t})$ into Eq. (S4), we have

$$(-\Gamma + i\omega)m_x = -m_y\left(\gamma N_y M_{eff} + i\alpha\omega - \alpha\Gamma\right)$$
$$+ \frac{\gamma}{M_S Vol}\tau_\parallel m_x \cot\theta - \frac{\gamma}{M_S Vol}\left[\frac{\partial \tau_\perp}{\partial V}\bigg|_\theta \frac{(50\Omega)}{R_0 + (50\Omega)} I \frac{\partial R}{\partial \theta}\bigg|_I - \frac{\partial \tau_\perp}{\partial \theta}\bigg|_V\right]m_y$$
$$(-\Gamma + i\omega)m_y = m_x\left(\gamma N_x M_{eff} + i\alpha\omega - \alpha\Gamma\right)$$
$$- \frac{\gamma}{M_S Vol}\left(\frac{\partial \tau_\parallel}{\partial V}\bigg|_\theta \frac{(50\Omega)}{R_0 + (50\Omega)} I \frac{\partial R}{\partial \theta}\bigg|_I - \frac{\partial \tau_\parallel}{\partial \theta}\bigg|_V\right)m_y - \frac{\gamma}{M_S Vol}\tau_\perp m_x \cot\theta.$$
(S21)

These equations determine the natural frequency $\omega$ and the decay rate $\Gamma$ of the damped resistance oscillation,

$$\omega \approx \gamma M_{eff}\sqrt{N_x\left[N_y - \frac{1}{M_{eff}M_S Vol}\left(\frac{\partial \tau_\perp}{\partial \theta}\bigg|_V + \frac{(50\ \Omega)}{R_0 + (50\ \Omega)} I \frac{\partial R}{\partial \theta}\bigg|_I \frac{\partial \tau_\perp}{\partial V}\bigg|_\theta\right)\right]}$$
(S22)

$$\Gamma \approx \frac{\alpha\gamma M_{eff}(N_x + N_y)}{2} - \frac{\gamma}{2 M_s Vol}\left(\tau_\parallel \cot\theta + \frac{\partial \tau_\parallel}{\partial \theta}\bigg|_V + \frac{(50\ \Omega)}{R_0 + (50\ \Omega)} I \frac{\partial R}{\partial \theta}\bigg|_I \frac{\partial \tau_\parallel}{\partial V}\bigg|_\theta\right).$$
(S23)

The frequency $\omega$ in Eq. (S22) is identical to the resonance frequency of ST-FMR in Eq. (S14), and the decay rate $\Gamma$ is indeed equal to $\Gamma_0$, the half linewidth of the ST-FMR resonance.

If we make further approximations in Eq. (S23) that $\frac{\partial \tau_\parallel}{\partial \theta}\bigg|_V \gg \frac{(50\ \Omega)}{R_0 + (50\ \Omega)} I \frac{\partial R}{\partial \theta}\bigg|_I \frac{\partial \tau_\parallel}{\partial V}\bigg|_\theta$ and use that for a tunnel junction $\tau_\parallel \propto \sin\theta$ (so that $\tau_\parallel \cot\theta = \partial \tau_\parallel/\partial \theta|_V$),[23] we have

$$\Gamma = \Gamma_0 \approx \frac{\alpha\gamma M_{eff}(N_x + N_y)}{2} - \frac{\gamma}{M_s Vol}\frac{\partial \tau_\parallel}{\partial \theta}\bigg|_V.$$
(S24)

The second term on the right differs by a factor of 2 from Eq. (A9) in ref. 9, on account of the correction we have made in the form of the LLGS equation [Eq. (S1) above], and it brings our result into accord with the calculations of Petit *et al.*[13,14]. This correction also improves the agreement between theory and experiment for the measurements for the bias dependence of the effective magnetic damping $\alpha_{eff}$ (defined as $\Gamma_0 = \alpha_{eff}\gamma M_{eff}(N_x + N_y)/2$), plotted in Fig. 7(a) of ref. 9. The correction does not affect the analysis of any of our other previous results to within experimental accuracy.

**Supplementary Note S4: Quantitative analysis of the decay rate of the resistance oscillations**

By our analysis in Supplementary Note S3 above, the decay rate $\Gamma$ measured from the time-domain resistance oscillations and the resonance half linewidth $\Gamma_0$ measured using DC-detected ST-FMR should be equal and can be expressed in terms of the effective damping $\alpha_{eff}$ of the free layer as $\Gamma = \Gamma_0 = \alpha_{eff} \gamma M_{eff} (N_x + N_y)/2$. Using $N_x \approx 4\pi$, $N_y \approx 0$, and $4\pi M_{eff} = 13 \pm 1$ kOe, we plot in Fig. S2 the bias dependence of the effective damping we determine from both $\Gamma$ (using the time-resolved measurements) and $\Gamma_0$ (using DC-detected ST-FMR). We observe quantitative agreement between the two methods within our experimental uncertainties. From this we conclude that the time-domain decay rate corresponds to the true decay of amplitude for the magnetization dynamics, with negligible contribution from dephasing between repetitions of the measurement. In the presence of significant dephasing between experimental repetitions, the time-domain oscillations would decay more quickly than the damping rate indicated by the ST-FMR linewidth.

From Eq. (S23), and taking into account that $\tau_\parallel \propto \sin\theta$, the effective damping should be related to the Gilbert damping and the spin transfer torque as:

$$\alpha_{eff} \approx \alpha - \frac{1}{M_s Vol M_{eff}(N_x + N_y)} \left( \left.\frac{\partial \tau_\parallel}{\partial \theta}\right|_V + \frac{(50\ \Omega)}{R_0 + (50\ \Omega)} I \left.\frac{\partial R}{\partial \theta}\right|_I \left.\frac{\partial \tau_\parallel}{\partial V}\right|_\theta \right) \qquad (S25)$$

The Gilbert damping coefficient $\alpha$ for the samples can be determined from the data in Fig. S2 near zero bias. We find $\alpha = 0.016 \pm 0.001$ for sample 2 (shown in Fig. S2) and $\alpha = 0.014 \pm 0.001$ for sample 1 (not shown). These values are at the high end of the typical reported value of Gilbert damping of CoFeB films,[40] possibly due to sidewall oxidation[41] or coupling to the SAF pinned layer. Using our measured spin torque and torkance results (Fig. 3 of the main paper), we can check whether the bias dependence of the damping that we measure is consistent with the prediction in Eq. (S25) (dashed lines in Fig. S2). We find good agreement, to within the level of uncertainty of the damping measurements. We should also note that the bias dependence of the effective damping could also be affected by factors such as heating and/or changes in the degree of spatial uniformity for the magnetic precession that are not included in our model.

**Supplementary Note 5: Calibration of the resistance of the MTJ and the applied bias voltage at high biases**

Calculation of the spin transfer torque based on Equations (2) and (3) of the main paper requires accurate measurements of the resistance of the MTJ (which depends on the applied magnetic field and bias voltage) and also the voltage provided by the square pulse during the time-domain ST-FMR measurement. Here we describe our calibration procedures for determining these quantities.

At low biases, the bias dependence of the differential resistance of the MTJ can be measured by a conventional low-frequency lock-in amplifier technique. However, this method is not practical at very high bias, because applying a large constant bias can damage or destroy the tunnel barrier even if the bias is still comfortably below the breakdown voltage for nanosecond pulses. In order to calibrate the resistance of the MTJ at high biases accurately, we utilize the fact that the amplitude of a reflected RF pulse from the MTJ depends on the differential resistance. For an incident RF pulse $V(t) = V_{in} \text{Re}(e^{i\omega t})$ with small amplitude $V_{in}$ having a frequency far away from resonance (so that it does not excite magnetic dynamics), the reflected amplitude is simply (from Eq. (S16))

$$V_{ref} = \frac{R_0 - (50\,\Omega)}{R_0 + (50\,\Omega)} V_{in}, \tag{S26}$$

where $R_0$ is the differential resistance, dependent on the offset angle and bias voltage across the MTJ. For calibration, we measure the reflected amplitude $V_{ref}$ using the same incident RF pulse for the whole range of different resistances we can access safely with a DC bias, ranging from a maximum value corresponding to the anti-parallel state with $V = 0$ to a minimum value corresponding to the parallel state with $V = 0.35$ V in our case. At the same time, we measure the differential resistances for all these states using a lock-in amplifier. This procedure provides the basis to convert $V_{ref}$ to $R_0$. We then apply square-wave pulses of various amplitudes (which can access higher biases) instead of a true DC voltage, and sweep the directions of the magnetic field (which changes the offset angle between the two moments), while measuring the reflected amplitude $V_{ref}$ of the RF pulse. Using this process, we map out $R_0$ as a function of offset angle and nominal amplitude $V_{sq}$ of the applied square pulse (read from the set output voltage of the pulse generator), i.e. $R_0 = R_0(\theta, V_{sq})$.

It should be noted that the calibration described so far has determined the differential resistance $R_0$ for any given square pulse (with nominal pulse amplitude $V_{sq}$), but the true value of bias voltage $V$ across the MTJ is not yet known. The equivalent bias $V$ under a square pulse depends on the DC resistance $R$:

$$V = \frac{2R}{R+(50\,\Omega)} \lambda V_{sq}, \quad (S27)$$

where $\lambda$ is a constant ratio between the nominal pulse amplitude and the true incoming pulse amplitude arriving on the device, determined by transmission and contact losses. In order to calibrate $V$ for any given $V_{sq}$, we must first determine the coefficient $\lambda$. Taking advantage of the significant bias dependence of resistance of the MTJ (especially in the anti-parallel state), this can be done easily within the DC-accessible range $|V| < 0.35$ V where both $R$ and $R_0$ can be directly measured by low-frequency techniques. We compare the value of $R_0$ as a function of $V_{sq}$ we measure from the microwave reflection method and a separate measurement of $R_0$ as a function of $V$ by low-frequency lock-in technique. This comparison provides a quantitative measurement of the bias $V$ on the MTJ for any $V_{sq}$ up to $|V| = 0.35$ V, which is sufficient to determine $\lambda$ using Eq. (S27). We have also further tested that Eq. (S27) with this calibrated $\lambda$ is indeed still accurate for higher bias (larger pulse height) by applying a large square pulse together with a DC voltage in the opposite polarity so that the added bias voltage is within the DC-accessible range and therefore can be calibrated.

After we determine $\lambda$, we must still determine the DC resistance $R$ in order to use Eq. (S27) to determine the true bias voltage $V$ at high bias. Since we have measured differential resistance $R_0 = R_0(\theta, V_{sq})$ using the RF pulse reflection method as discussed above, $V$, $I$ and $R$ can all be determined by numerical integration from low bias to higher bias. Because $R=V/I$, we can write Eq. (S27) as $2\lambda V_{sq} = V + I(50\,\Omega)$, which can then be used to determine the relationship between an incremental change in incident pulse height and the incremental change in voltage or current that it produces:

$$dV = 2\lambda dV_{sq} \left/ \left[1 + \frac{(50\,\Omega)}{R_0}\right] \right. \quad \text{or} \quad dI = 2\lambda dV_{sq} / [R_0 + (50\,\Omega)], \quad (S28)$$

where $R_0 = dV/dI$ is the differential resistance. Therefore,

$$V(\theta,V_{sq}) = \int_0^{V_{sq}} \frac{2\lambda R_0(\theta,V_{sq})}{R_0(\theta,V_{sq}) + (50\ \Omega)} dV_{sq}$$

$$I(\theta,V_{sq}) = \int_0^{V_{sq}} \frac{2\lambda}{R_0(\theta,V_{sq}) + (50\ \Omega)} dV_{sq} \qquad (S29)$$

$$R(\theta,V_{sq}) = \frac{V(\theta,V_{sq})}{I(\theta,V_{sq})}$$

We can thus obtain calibrated values of $V$, $R$ and $R_0$ at any given $\theta$ and $V_{sq}$, as required to calculate the spin-transfer torque using Equations (2) and (3) of the main paper.

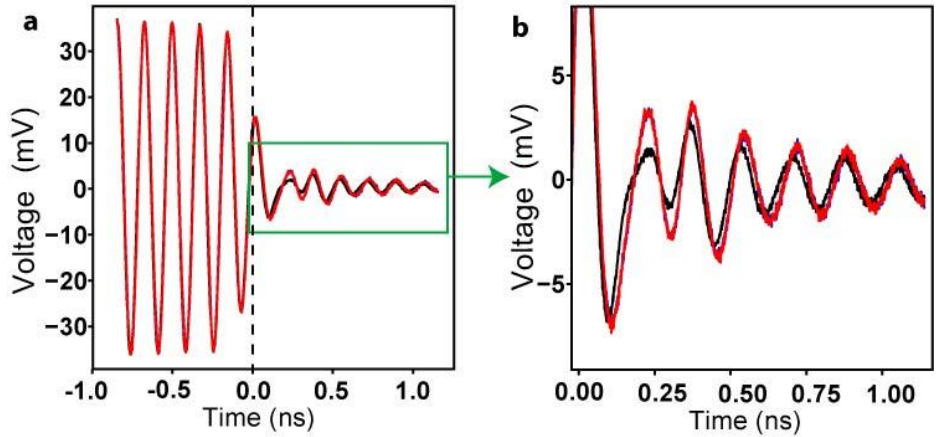

**Supplementary Figure S1: Background subtraction. a**, The raw signal and background waveform traces used to measure the resistance oscillation signal in Fig. 2 of the main paper. The red curve and the blue curve (overlapping with the red curve and hardly visible) are the reflected voltage waveforms measured by the sampling oscilloscope for a 5.8 GHz RF driving pulse, taken from sample 1 under the following bias parameters: magnetic field $H$ = 200 Oe, field direction $\beta$ = 90°, offset angle $\theta$ = 85°, $V$ = 0.38 V. The black curve is the reflected voltage waveform measured for the same driving pulse but for a different magnetic field $H$ = 600 Oe, $\beta$ = 94°, so as to provide a non-resonant background measurement. **b**, Zoom-in of the curves within the green box in **a**. Subtraction of the background curve from the measured waveform near resonance yields the result in Fig. 2b and 2c of the main paper.

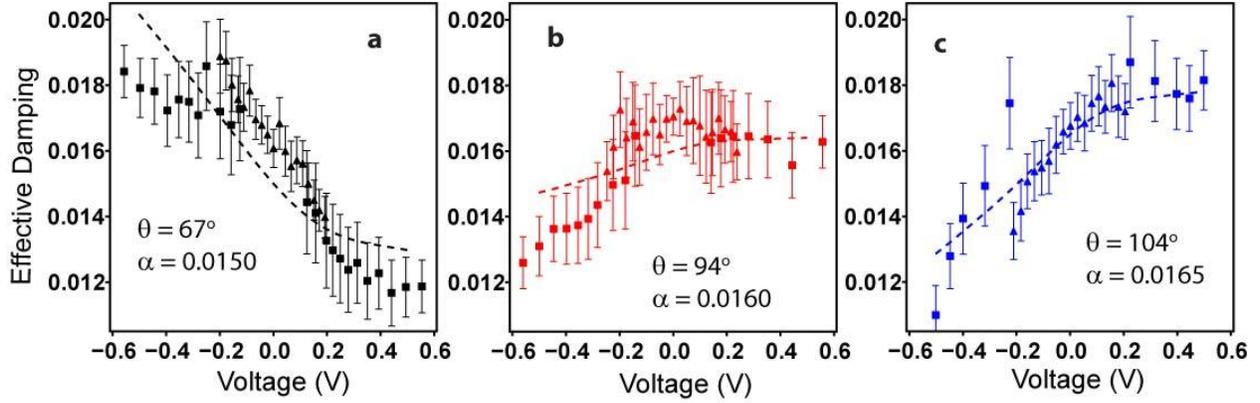

**Supplementary Figure S2: Measurement of effective damping.** (squares) Bias dependence of the effective damping $\alpha_{eff}$ determined from the decay rate $\Gamma$ of the resistance oscillation in the time-resolved measurements (averaging over RF driving frequencies close to the resonance peak). (triangles) Bias dependence of the effective damping determined from the linewidth $\Gamma_0$ of the resonance peak in DC-detected ST-FMR measurements. The different panels correspond to different initial offset angles $\theta$ for the same sample (sample 2). The dashed lines show the effective damping predicted by Eq. (S25), using the in-plane torkances measured in the time-resolved experiment. Slightly different values of the Gilbert damping $\alpha$ were used for different offset angles to provide better fits of the data.